# Investigating the Necessity of Distribution Markets in Accomodating High Penetration Microgrids

Sina Parhizi and Amin Khodaei
Department of Electrical and Computer Engineering
University of Denver
Denver, Colorado
sina.parhizi@du.edu, amin.khodaei@du.edu

*Abstract*— **The increased need for reliable, resilient, and high quality power combined with a falling cost of distributed generation technologies has resulted in a rapid growth of microgrid in power systems. Although providing multitude of benefits, the microgrid power transfer with the main grid, which is commonly obtained using economy and reliability consideration, may result in major operational drawbacks, most notably, a large mismatch between actual and forecasted system loads. This paper investigates the impact of high penetration microgrids on the power system net load, and further proposes three paradigms that can be adopted to address the emerging operational issues. The IEEE 6-bus test system is used for numerical studies and to further support the discussions.**

## Nomenclature

| | |
|---|---|
| $D$ | Load demand. |
| $i$ | Index for DERs. |
| $I$ | Commitment state of dispatchable unit (1 when committed, 0 otherwise). |
| $F(P,I)$ | Dispatchable unit operation cost. |
| $LS$ | Load curtailment. |
| $m$ | Index for microgrids. |
| $P$ | DER output power. |
| $P^M$ | Power transfer to the microgrid. |
| $t$ | Index for hours. |
| $\rho^M$ | Market price. |
| $v$ | Penalty for scheduled power violation. |
| $\upsilon$ | Value of lost load. |

## I. Introduction

**M**ICROGRIDS, as groups of interconnected loads and distributed energy resources (DERs) with clearly defined have electrical boundaries and the capability to operate in the grid-connected and islanded modes, were primarily introduced to facilitate the integration of high penetration DERs to distribution grids [1]. Microgrids offer several benefits to consumers and the system as a whole including improved reliability and resiliency, local intelligence to the customer side, reduction in greenhouse gas emissions, and reducing the need for expanding transmission and distribution facilities as a result of generation-load proximity [2]-[10]. There has been significant investment on microgrids during the past decade, where it is estimated that the installed microgrid capacity would grow from the 1.1 GW in 2012 to 4.7 GW in 2017 with an estimated market opportunity of $17.3 billion [11].

The increased controllability of microgrids, however, is perceive both as an opportunity and as a challenge for the main power system. On one hand, microgrids can take part in efficient DER and demand response integration and to ensure a widespread adoption of these new technologies which would help the system in reaching economic objectives and meeting environmental mandates [12]. On the other hand, microgrids commonly rely on price-based schemes to manage local DERs and loads which may result in a highly uncertain and variable net load. Under the price-based scheme, the microgrid objective is to minimize its operation cost (considering the cost of local generation and cost of energy purchase from the main grid) while taking security issues into account. In this case, when the electricity price in the main grid is low the microgrid would purchase power from the main grid and reduce its local generation (i.e., a large positive net load), however when the electricity price is high the microgrid would prefer to sell excess generation back to the main grid to increase its economic benefits (i.e., a large negative net load). As price varies during the day, the microgrid DERs and loads schedules will change to ensure the minimum cost and maximum savings. These changes will result in an increased level of load uncertainty in the system, which would accordingly challenge a reliable supply-demand balance, increase day-ahead load variations and the need for load following and frequency regulation services, and result in a sub-optimal resource scheduling solution obtained by the system operator. These issues will undoubtedly be more noticeable as the microgrid penetration increases in distribution networks.

Challenges in integration of microgrids and responsive loads have triggered many efforts to establish markets at the distribution level [13]. One of the primary objectives of distribution markets is to shift the microgrid scheduling from price-based schemes to market-based schemes, and accordingly, resolve the challenges in high penetration microgrid deployment by reducing net load variability and uncertainty. Currently, a few distribution market models are under investigation in the United States. In [14] a price-based simultaneous operation of microgrids and a Distribution Network Operator (DNO) is proposed. In New York, a new entity, called Distributed System Platform Provider (DSPP), is introduced via the Reforming the Energy Vision program [15]. The DSPP can establish a universal market environment instead of one for each utility. In California, the state public utilities commission has ruled to establish regulations to guide investor-owned electric utilities in developing their Distribution Resources Plan proposals. Studies in [16] and [17] provide a framework for this ruling and define an entity referred to as Distribution System Operator (DSO), to be in charge of operation of local distribution area and providing distribution services. It would be responsible for forecasting and measurement services to the ISO and managing the power flow across the distribution system. The study in [18] proposes the DSO to be an ISO for the distribution network, responsible for balancing supply and demand at the distribution level, linking wholesale and retail market agents, and linking the ISO to the demand side. It describes a spectrum of different levels for DSO autonomy in operating the distribution system and the degree of ISO's control over it. From the least autonomy to the most autonomy, this spectrum entails DSO to be able to perform the forecasting and send it to the ISO, be responsible for balancing the supply and demand, be able to receive offers from DER units, aggregate them and bid it into the wholesale market, and eventually be able to control the retail market so that different DERs can have transactions not only with the DSO but among themselves. In [19], an independent distribution system operator (IDSO) is proposed to be responsible for distribution grid operation, while grid ownership remains in the hands of utilities. The IDSO is envisioned to provide market mechanisms in the distribution system, enable open access, and ensure safe and reliable electricity services. The IDSO will reduce the operation burden on utilities and determine the true value of resources more objectively.

This paper investigates the impact of price-based microgrid scheduling schemes on the main power system with respect to price variability and load uncertainty for various levels of microgrid penetration. The results obtained from the proposed models will clearly demonstrate the necessity of implementing distribution markets to manage high penetration microgrids. The mixed integer linear programming is used to model the microgrid optimal scheduling problem, where it is further tested on the IEEE 6-bus test system.

The rest of the paper is organized as follows. Section II formulates the price-based microgrid scheduling problem, Section III presents the obtained results from testing the IEEE 6-bus system, and Section IV provides the conclusions.

## II. PRICE-BASED OPTIMAL SCHEDULING MODEL

Microgrid control is commonly performed in a hierarchical three-level scheme, including primary, secondary, and tertiary levels [20], [21]. The first two levels are responsible for droop control and frequency/voltage regulations in response to load variations and/or islanding. At the third level, the microgrid controller seeks to minimize the microgrid operation cost, i.e., the generation cost of local DERs, as well as the energy exchange with the main grid, to supply forecasted local loads in a certain period of time (typically one day). This problem is subject to a variety of operational constraints, such as power balance and DER limitations. The scheduling problem can be solved centrally through a central controller [22], [23] or in a decentralized way where each entity communicates with others as an agent to obtain the optimal schedule for the entire microgrid [24], [25]. A variety of methodologies are proposed in the literature to solve the microgrid optimal scheduling problem, including deterministic, heuristic, and stochastic methods. Mixed integer programming (MIP) is widely used to formulate resource scheduling problems [26]-[28] and is further used here to model the microgrid price-based scheduling problem.

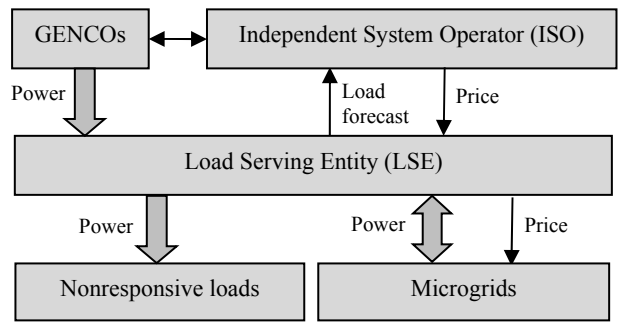

Fig. 1. The price-based microgrid scheduling framework.

The wholesale market structure considering the microgrid price-based scheduling scheme is shown in Fig. 1. The aggregate demand of the distribution network, including the demand of microgrids, responsive customers, and nonresponsive customers, is forecasted and submitted to the ISO by the load serving entity (LSE). The LSE is the utility company which is responsible for ensuring a reliable flow of power from generation companies to customers via transmission and distribution networks. Based on the demand data, as well as the generation and transmission data obtained respectively from generation companies (GENCOs) and transmission companies (TRANSCOs), the ISO runs the unit commitment and economic dispatch problems to determine the optimal schedule of generation units and the locational marginal prices (LMPs) at every system bus. Microgrids use LMPs at their associated bus and solve the price-based optimal scheduling problem as defined in (1)-(6):

$$\min \sum_t \sum_i (F_{im}(P_{imt}, I_{imt}) + \upsilon LS_{mt} + \rho_m^M P_{mt}^M) \quad (1)$$

$$P_{mt}^M + \sum_i P_{imt} + LS_{mt} = \sum_d D_{mdt} \quad \forall t \quad (2)$$

$$P_{im}^{\min} I_{imt} \leq P_{imt} \leq P_{im}^{\max} I_{imt} \quad \forall t, \forall i \quad (3)$$

$$\sum_t P_{imt} = E_{im} \quad \forall t \quad (4)$$

$$\sum_t f_{im}(P_{imt}, I_{imt}) \leq 0 \quad \forall i \quad (5)$$

$$-P_m^{M,\max} \leq P_{mt}^M \leq P_m^{M,\max} \quad \forall t \quad (6)$$

The three terms in the objective function represent the operation cost, the load curtailment cost, and the main grid power transfer cost. The operation cost is the cost of power production by dispatchable units as well as startup and shut down costs. The load curtailment cost is defined as the value of lost load times the amount of load curtailment. The value of lost load is assumed as on opportunity cost based on the cost the consumer is willing to pay to have reliable uninterrupted service. It is commonly used as a measure to represent loads criticality [29]. The power transfer cost is equal to the amount of power transferred to the microgrid from the main grid times the associated LMP to which the microgrid is connected. The objective is subject to a set of operational constraints. The power balance constraint (2) ensures that the sum of the main grid power transfer plus the locally generated power matches the microgrid load, while load curtailment variable is added to ensure that this balance is satisfied at all times (in particular during the islanded operation when adequate generation may not be available). In the power balance equation, nondispatchable unit generation and fixed load values are forecasted, where dispatchable unit generation, adjustable load, load curtailment, and energy storage power are the variables. All operational constraints associated with DERs and loads are formulated using three general constraints (3)-(5), respectively representing power constraints, energy constraints, and time-coupling constraints. Power constraints (3) account for generation minimum/maximum capacity limits, storage minimum/ maximum charge/discharge power, and flexible load minimum/maximum capacity limits. Energy constraints (4) account for energy storage state of charge limit and flexible load required energy in each cycle. Time-coupling constraints (5) represent any constraint that link variables in two or more scheduling hours, including dispatchable units ramp up/down, minimum on/off times, energy storage rate and profile of charge/discharge, and adjustable loads minimum operating time and load pickup/drop rates. The detailed formulation of these constraints can be found in [4]. The main grid power transfer is restricted by its associated limits, which are imposed by the capacity of the line connecting the microgrid to the main grid, in (6).

## III. ILLUSTRATIVE EXAMPLE

The IEEE 6-bus test system is used to demonstrate the impact of the price-based microgrid scheduling on changing the system net load. The IEEE 6-bus system data and the microgrid data are borrowed from [30] and [4], respectively. Fig. 2 shows the aggregated system net load with 50% microgrid penetration at each load bus in two cases: i) the forecasted load that is provided by the utility to the ISO. The ISO has used the forecasted load to determine the commitment and the dispatch of available generation units, and further calculate the LMPs; and ii) the actual load once calculated LMPs are sent to microgrids and microgrids have scheduled their DERs and loads. As this figure demonstrates, microgrids can potentially result in a complete change in the system load profile. The ISO's challenge is to remove the mismatch between the generation and the load, since generation units are committed and dispatched based on the forecasted load while the system encounters a revised load.

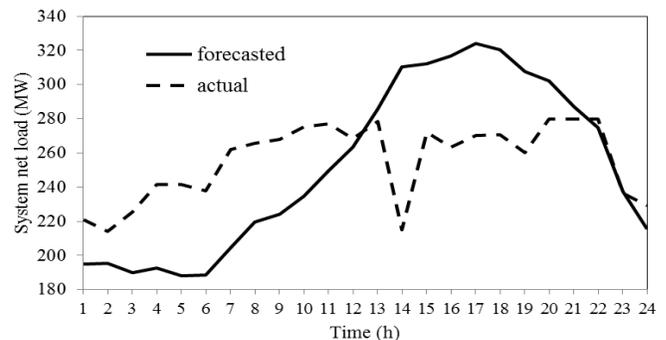

Fig 2. The aggregated system load in two cases: forecasted load by the utilities, and the actual load after microgrids scheduling.

## IV. PROPOSED PARADIGMS

This paper discusses three paradigms to address the load uncertainty challenges introduced by high penetration microgrids: 1) Removing the generation-load imbalance by redispatching committed generation units, i.e., similar to the current practice in grid control; 2) Communicating the revised load to the ISO to solve the unit commitment again and obtain new unit schedule and LMP values; 3) Introducing a distribution market to locally manage microgrids, i.e., to shift from the price-based scheduling scheme to a market-based scheduling scheme.

### A. Paradigm 1

The ISO would redispatch the committed generation units to compensate the mismatch created due to the change in system demand. Fig. 3 shows the required change in generation at each bus (the other generator dispatch did not change). Under this paradigm, the generation-load imbalance can be eliminated; the amount of change in dispatch is significant and might not be feasible in some occasions without a change in unit commitment.

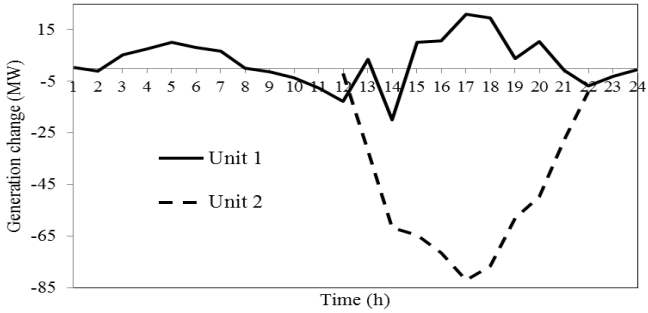
Fig. 3. Change in generation after ISO redispatches the units.

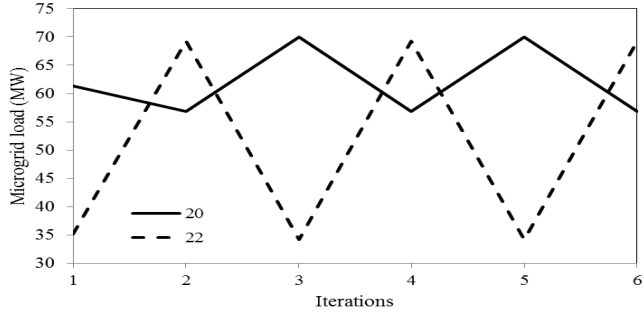
Fig. 4. Microgrid load at hours 20 and 22.

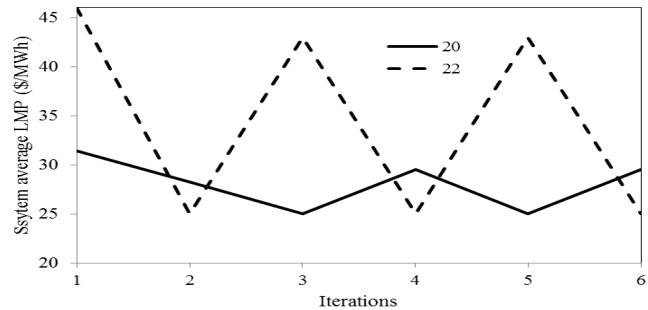
Fig.5. System average LMP at hours 20 and 22.

### B. Paradigm 2

Under this paradigm, the revised load is communicated to the ISO to solve the unit commitment again and obtain new unit schedule and LMP values. This paradigm should be performed in a day-ahead fashion and would require a reliable communication infrastructure among the ISO and microgrids. An example for this paradigm is shown in Fig. 4, where it is assumed that load at bus 3 can be fully supplied by local microgrids. The microgrid load is oscillating as the ISO and microgrids communicate price and load in each iteration.

Fig. 5. Shows the system average LMP at the same hours. When the microgrid responds to the price set by the ISO, the ISO has to reschedule the system resources resulting in a new price. Microgrids will also reschedule their resources according to this new price. As the number of buses with microgrids increases, the price oscillations also increase. With a larger penetration of microgrids in the system, the system load becomes more responsive and causes more volatility in system LMPs.

### C. Paradigm 3

The uncertainty in the system operation and the need to commit adequate reserve to support load variations causes troubles for the ISO to reliably operate the system. Hence, alternative models to manage microgrids are being actively sought. Under this paradigm a new entity, here called Distribution Market Operator (DMO), is introduced to establish a competitive electricity market in the distribution level. Microgrids would be players in the distribution market and participate in the electricity price calculations. Microgrids would submit their demand bids to the DMO, which would in turn aggregate the bids and submit it to the ISO. The ISO determines the awarded power to each DMO, where the DMO subsequently disaggregate the awarded bids to participated microgrids. Microgrids would be obliged to follow the awarded power once the market is cleared. This would significantly reduce uncertainties the ISO faces as the penetration of microgrids in the system increases. The proposed market-based scheduling model is depicted in Fig. 6.

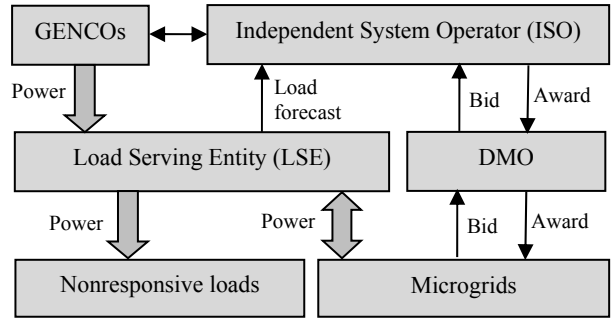
Fig. 6. Proposed microgrid market participation through he DMO.

The proposed framework offers several advantages:
- The microgrid demand is set by the DMO and known with certainty on a day-ahead basis. This will lead to manageable peak demands and increased operational reliability and efficiency.
- The microgrid can exchange power with the main grid and act as a player in the electricity market. The DMO would serve as an interface between the ISO and microgrids that facilitates microgrids market participation and coordinates the microgrids with the main grid to minimize the risks posed by microgrids operational uncertainties.
- Establishment of the DMO is beneficial to the ISO as it allows a significant reduction in the required communication infrastructure among microgrids and the ISO.
- The DMO can be formed as a new entity or be part of the currently existing electric utilities. An independent DMO would be able to set up a universal market environment instead of one for each utility. It would also be less suspected of exercising market power. On the other hand, a utility-affiliated DMO would be able to perform several functionalities currently possessed by electric utilities without necessitating additional investments.

Considering the listed advantages, and many more that will be obtained using more detailed numerical simulations in future work, distribution markets can be considered as both beneficial and necessary components in modern power grids which will help accommodate a large penetration of active customers.

## V. CONCLUSION

The microgrid price-based scheduling may cause load and price oscillations in the system since there is a high probability that microgrids follow a different schedule compared to the one forecasted by the utility once actual prices are determined. The increase in the number of entities with responsive loads operated based on price-based schemes, and in particular microgrids, would intensify this issue. In other words, setting the price centrally by the system operator and sending it to microgrids, so they can accordingly schedule their resources, can potentially result in significant uncertainty in the system. This paper provides a study of this phenomenon in price-based scheduled microgrids integrated in the power system. In order to manage this issue, market-based scheduling models can be pursued to allow microgrids become players in a distribution market and reduce the uncertainty in microgrid demand from the ISO perspective.